\documentstyle[aps,preprint]{revtex}
\begin{document}
\draft
\preprint{SOGANG-HEP-206/96}
\title{Entropy of the BTZ Black Hole in 2+1 Dimensions}
\author{Sung-Won Kim\footnote{Electronic address: sungwon@mm.ewha.ac.kr}}
\address{
Department of Science Education and Basic Science Research Institute,\\
Ewha Women's University, Seoul 120-750, Korea}
\author{Won T. Kim\footnote{Electronic address: wtkim@physics.sogang.ac.kr},
    Young-Jai Park\footnote{Electronic address: yjpark@physics.sogang.ac.kr},
and Hyeonjoon Shin\footnote{Electronic address: hshin@physics.sogang.ac.kr} }
\address{
Department of Physics and Basic Science Research Institute, \\ 
Sogang University, Seoul 121-742, Korea }
\date{April 1996, revised version}
\maketitle
\begin{abstract}
We semi-classically calculate the entropy of a scalar field in the
background of the BTZ black hole, and derive the perimeter law of the
entropy. The proper length from the horizon to the ultraviolet
cutoff is independent of both the mass and the angular momentum of the
black hole. It is shown that the superradiant scattering
modes give the sub-leading order contribution to the entropy while the
non-superradiant modes give the leading order one, and thus superradiant
effect is minor.
\end{abstract}
\bigskip

\newpage
Two decades ago, Bekenstein suggested that the entropy of a
black hole is proportional to the area of the horizon through
the thermodynamic analogy \cite{bek}.
Subsequently, Hawking showed that the entropy of the Schwarzschild black hole
satisfies exactly the area law by means of Hawking radiation based on
the quantum field theory \cite{haw}.
On the other hand, 't Hooft has argued that when one calculates the
black hole entropy, the modes of a quantum field in the vicinity of a 
black hole horizon should be cut off due to gravitational effects
rather than infinitely piling up by imposing a brick-wall cutoff
just beyond the horizon.
Nowadays the evaluation of black hole entropy in terms of statistical
mechanics is one of the outstanding quantum gravity problems in connection
with the information loss problem \cite{sus,kab}.

On the other hand, there has been much interest
in lower dimensional theories of
gravitation with the aim of the consistent quantum gravity.
In a (1+1)-dimensional gravity, Callan-Giddings-Harvey-Strominger (CGHS)
model \cite{cal}, which has been improved by Russo, Susskind, and 
Thorlacius (RST), gives the analytic solution of an evaporating black hole in
the semi-classical approximation \cite{rus}. Recent studies on the black hole
thermodynamics based on these models \cite{mye,hay,Rus}
show that the entropy
satisfies the area law and thermodynamic second law.
For the (2+1)-dimensional anti-de Sitter gravity,
Ba$\tilde{{\rm n}}$ados, Teitelboim, and Zanelli
(BTZ) have obtained the black hole solution, which is asymptotically
anti-de Sitter rather than asymptotically flat
and is characterized by the mass and the angular
momentum \cite{ban}.
For the consistent formation of the event horizon, the angular
momentum should be restricted to some values.
The various thermodynamic properties of the BTZ black hole were shown in
Refs. \cite{brow,zasl}.
Similar to the (1+1)-dimensional dilation gravity \cite{kim},
there is no dynamically propagating degrees of freedom
in contrast to the four dimensional Einstein gravity.
Thus the BTZ black hole can also be a good candidate in studying the quantum
aspects of black holes without the complexity of degrees of freedom.

Once the BTZ black hole is assumed, it is natural to consider some particles
or fields around the black hole due to the quantum fluctuation of particles
so called Hawking radiation. Therefore one can consider
the black hole-matter
coupled action as a total system.
It would be interesting to confirm the area law for the
BTZ black hole since it can be an universal property of
black holes.
As well-known, the gravitational part of the
entropy \cite{gib} was already
calculated in \cite{ban,cali,Cali}, and the area law is satisfied.
However, we could not assert generally the area law
of black holes, such that in Ref. \cite{ich}
the entropy of matters on the BTZ black hole
does not seem to satisfy the perimeter law.
Moreover, the geometrical structure of the BTZ black hole
in 2+1 dimensions is somewhat
different from the usual Schwarzschild black hole (or Kerr black
hole) in four dimensions.

In this paper, we shall recast the entropy of a matter on
the BTZ black hole background. By using the brick wall method
developed by 't Hooft \cite{tho}, we shall consider a Klein-Gordon field
on the BTZ black hole, and obtain the free energy in the semi-classical
approximation. 
We shall consider the entropy for the non-rotating and the rotating case
separately. For the non-rotating black hole, we obtain the free
energy which gives the desired entropy formula.
For the rotating case, the free energy is composed of two pieces, 
one is the non-superradiant (NSR) part and the other is the superradiant 
(SR) part. It is shown that the NSR part gives the leading order contribution 
to the free energy in terms of the brick wall cutoff while the SR part gives 
the sub-leading order one. It is also shown that, although the SR part
contains a  
divergent term due to the large angular quantum number, which is absent in 
the NSR part, it does not contribute to the entropy.
After all, for both cases, we obtain the entropy expressed
in terms of the perimeter by choosing the intrinsic ultraviolet cutoff, 
which is independent of mass and angular momentum of the black hole.
This result is similar to the four dimensional case obtained by 't Hooft.

Let us now introduce
the (2+1)-dimensional gravity, which is given by
\begin{equation}
\label{action}
I=\frac{1}{2\pi} \int d^3 x \sqrt{-g} \left[R +\frac{2}{l^2} \right] +B~,
\end{equation}
where $\Lambda= -\frac{1}{l^2}$ is the cosmological constant and $B$ is 
the boundary term.
Then the classical equation of motion yields the BTZ metric as
\begin{eqnarray}
\label{metric}
ds^2 &=& g_{tt} dt^2 - J dt d\theta
 + \frac{1}{N^2(r)} dr^2 + r^2 d \theta^2, \\
g_{tt} &=&- \left( \frac{r^2}{l^2}-M \right), \\
N^2(r) &=& \frac{r^2}{l^2} - M + \frac{J^2}{4 r^2}. \nonumber
\end{eqnarray}
There exist two coordinate singularities corresponding to
the outer and inner horizon from $N^2(r) = 0$,
\begin{equation}
\label{horizon}
r_\pm =  \sqrt{M} l\left[ \frac{1}{2} \left( 1 \pm
   \sqrt{1- \left(\frac{J}{Ml}\right)^2 } \right) \right]^{1/2}~~
   (~|J| \leq Ml~).
\end{equation}

>From now on, we shall consider only the nonextremal case, $r_+ \neq r_-$.
We regard the black hole horizon as $r_H = r_+$, which has the non-rotating
limit.
In later convenient use, some quantities are rewritten by $r_\pm$,
\begin{eqnarray}
N^2(r) &=& \frac{1}{l^2 r^2} (r^2-r^2_+)(r^2-r_-^2), \\
M &=& \frac{r_+^2 + r_-^2}{l^2}, \\
\label{J}
J &=& \frac{2 r_+ r_-}{l}.
\end{eqnarray}
The stationary limit $r_{\rm erg}$ which is defined by
the radius of the ergosphere, is
obtained by solving $g_{tt} = 0$ as follows
\begin{equation}
\label{erg}
r_{\rm erg} = \sqrt{M}l = \sqrt{r_+^2 + r_-^2}.
\end{equation}
The angular velocity of the black hole horizon is defined by
\begin{equation}
\label{angular}
\Omega_H = - \left. \frac{g_{t \theta}}{g_{\theta \theta} }
   	     \right|_{r = r_+}
         = \frac{J}{2 r_+^2} = \frac{r_-}{l r_+}.
\end{equation}

Let us now introduce a Klein-Gordon field equation on the BTZ black hole
background,
\begin{equation}
\label{kg}
\frac{1}{\sqrt{-g}} \partial_\mu ( \sqrt{-g} g^{\mu \nu}
  \partial_\nu \Phi ) - \mu^2 \Phi = 0,
\end{equation}
where $\mu$ is the mass of a scalar field $\Phi$.
The above equation can be solved through the separation of variables,
i.e., we can write the wave function as
\begin{equation}
\label{wf}
\Phi(r, \phi, t) = e^{-iEt}e^{i m \phi} R_{Em} (r),
\end{equation}
where $m$ is the azimuthal quantum number.
Then, the radial equation becomes
\begin{equation}
\label{eq11}
\frac{1}{r} \partial_r [ r N^2(r) \partial_r R_{Em}(r)] +
N^2(r) k^2 (r,m,E) R_{Em}(r) = 0~,
\end{equation}
where the $r$-dependent radial wave number is given by
\begin{equation}
\label{wn}
k^2 (r,m,E) = \frac{1}{N^4(r)} \left[ E^2 - \mu^2 N^2(r) -
  \frac{JEm}{r^2} + \frac{m^2 (M-r^2/l^2)}{r^2} \right]
\end{equation}
in the WKB approximation \cite{tho,man}.
According to the semi-classical quantization rule, the
radial wave number is quantized as
\begin{equation}
\label{rule}
\pi n_r = \int^L_{r_H+ \epsilon} dr k(r,m,E)~,
\end{equation}
where $n_r$ is assumed to be a nonnegative integer,
and $\epsilon$ and $L$ are ultraviolet and infrared regulators,
respectively. 
This is nothing but the quantization condition of energy since
$E=E(n_r,m)$ by inverting the relation (\ref{rule}).
The number of modes with energy not exceeding $E$ is obtained
by assuming the first brick wall (ultraviolet regulator) to be
located at just outside of the outer horizon. 
Accordingly, we consider the radial integration for 
$r > r_{\rm H}$.
On the other hand, for sufficiently large $r$ the radial wave vector
$k$ can be a complex value unless we put the infrared regulator as
$r_{\max} =L \approx \frac{1}{\mu}$. In later calculation
for the free energy, we shall set $L \rightarrow \infty$ for the massless
scalar particle $\mu^2 =0$ for simplicity without losing consistency.

The free energy at inverse temperature $\beta$
on the rotating black hole with the angular
velocity $\Omega_H$ is represented by \cite{ich,thor}
\begin{equation}
\label{def}
e^{-\beta F} = \prod_K
                \left[ 1 - e^{-\beta(E_K-m\Omega_H)} \right]^{-1}~,
\end{equation}
where $K$ represents the set of quantum numbers.
Note that the dependence on $E_K -m\Omega_H$ rather
than $E_K$ is a sign that the BTZ black hole
has superradiant scattering modes for the rotating case.
By using Eq. (\ref{rule}), the free energy can be rewritten as
\begin{eqnarray}
 F &=& \frac{1}{\beta}\sum_K \ln \left[ 1 - e^{-\beta(E_K-m\Omega_H)} \right]
   ~\approx ~\frac{1}{\beta} \int dn_r \int dm ~\ln 
            \left[ 1 - e^{-\beta(E-m\Omega_H)} \right]
            \nonumber   \\
   &=& -\int dm \int dE ~\frac{n_r}{e^{\beta (E-m\Omega_H)} -1}
            \nonumber  \\
\label{free}
   &=& -\frac{1}{\pi} \int dm \int dE
         \frac{1}{e^{\beta (E-m\Omega_H)} -1}
         \int^L_{r_H+\epsilon} dr k(r,m,E)~,
\end{eqnarray}
where we have taken the continuum limit in the first line and integrated
by parts in the second line in Eq. (\ref{free}).

We first consider the non-rotating black hole. Since it does not
rotate, $\Omega_H=0$, $J=0$, or $r_-=0$ as can be seen from
Eq. (\ref{angular}). The free energy (\ref{free}) then becomes
\begin{equation}
\label{free00}
 F_{(J =0)}  = -\frac{1}{\pi} \int dm \int^\infty_0 dE
         \frac{1}{e^{\beta E} -1}
         \int^L_{r_H+\epsilon} dr k(r,m,E)~.
\end{equation}
For the evaluation of the physical free energy, the guideline provided
by the brick wall method is to make the result of integration be real.
Following this line, if we perform $m$-integration of
Eq. (\ref{free00}) firstly and the remaining radial and energy
integrations, then the form of the free energy is simply
obtained as follows
\begin{eqnarray}
F_{(J=0)} &=& -\frac{1}{2} \int^\infty_0 dE
      \frac{1}{e^{\beta E}-1} \int^L_{r_H +\epsilon} dr 
       \frac{l r}{(r^2-r_+^2)^{3/2}} 
     \left[ l^2 E^2 - \mu^2 (r^2-r_+^2) \right] \nonumber \\
  &=& \frac{l\mu^2}{2\beta}
       \int^\infty_0 dz \frac{1}{e^z-1}   
       ( L - \sqrt{(r_H +\epsilon)^2-r_+^2}) \nonumber \\
   \label{free0}
  & & + \frac{\zeta(3) l^3}{\beta^3}
        \left( \frac{1}{L} -
               \frac{1}{\sqrt{(r_H +\epsilon)^2-r_+^2} }
        \right)~,
\end{eqnarray}
where  $z \equiv \beta E$.
Now, if the scalar field is massless, i.e., $\mu = 0$, for simplicity,
and the limit
$L \rightarrow \infty$ is taken, the free energy (\ref{free0}) is 
simplified as follows
\begin{equation}
\label{freeJ0}
F_{(J=0)} =  - \frac{\zeta(3) l^3}{\beta^3}
             \frac{1}{\sqrt{(r_H +\epsilon)^2-r_+^2} }~,
\end{equation}
which is exact in the sense of WKB approximation.

Let us now turn to the evaluation of the entropy for the massless
field, which can be obtained from the
free energy (\ref{freeJ0}) at the black hole temperature.
Since we are considering the non-rotating black hole and will also deal
with the rotating one later, we designate the entropy for the
non-rotating case as $S_{(J=0)}$, while for the rotating case as 
$S_{(J\neq0)}$.
For the non-rotating case, the inverse of Hawking temperature, 
$\beta_H$, is given by
\begin{equation}
\label{betaJ0}
\beta_H = \frac{2\pi l^2}{r_+},
\end{equation}
then the entropy is
\begin{eqnarray}
S_{(J=0)} &=& \left. \beta^2 \frac{\partial F_{(J=0)}}{\partial \beta}
              \right|_{\beta=\beta_H}
                                \nonumber \\
\label{entropy0}
  &=&  \frac{3 \zeta(3) l^3}{\beta_H^2} 
       \frac{1}{\sqrt{(r_H+\epsilon)^2-r_+^2} }~~.
\end{eqnarray}
This result clearly shows that the entropy behaves like
$1/\sqrt{\epsilon}$ as $\epsilon \rightarrow 0$, that is, it is
divergent in terms of the ultraviolet brick wall cutoff $\epsilon$. 
At this level, $\epsilon$ is not determined. 
However, by requiring that the entropy $S_{(J=0)}$ satisfies a
statement about entropy, i.e. the area law, the brick wall cutoff 
$\epsilon$ may be determined as a finite value. The process of
determining $\epsilon$ may be reversed: choosing $\epsilon$ making
$S_{(J=0)}$ satisfy the area law.
If we now choose the cutoff as
\begin{equation}
\label{epsilonJ0}
\epsilon = r_+ \left( \sqrt{1+\frac{a^2}{l^2} } -1 \right)~,
\end{equation}
where the constant $a$ is defined by
\begin{equation}
\label{adef}
a \equiv \frac{3 \zeta(3)}{2^4 \pi^3}
\end{equation}
and its numerical value is approximately 
$a \approx 7.3 \times 10^{-3}$,
then the entropy (\ref{entropy0}) satisfies the area (perimeter) law,
\begin{equation}
\label{area}
S = 2 \cdot 2 \pi r_+.
\end{equation}

The cutoff (\ref{epsilonJ0}) seems to depend on the mass of the 
black hole. But the invariant cutoff will be independent of it
as far as we require the perimeter law of the entropy.
The invariant distance from the horizon to the brick wall
is calculated by definition as
\begin{eqnarray}
\label{invariant}
\tilde{\epsilon}&=&\int^{r_H +\epsilon}_{r_+}\frac{1}{N(r)} dr \\ \nonumber
                &=& l \ln \left(\frac{\sqrt{(r_H+\epsilon)^2 -r_+^2}
                +(r_H+\epsilon)}{r_+}\right).
\end{eqnarray}
After some calculations, the entropy (\ref{entropy0})
is neatly represented in terms of
the invariant cutoff (\ref{invariant}) as follows,
\begin{equation}
S_{(J=0)} =\frac{4\pi a}{l} \frac{r_+}
{\sinh \left( \frac{\tilde{\epsilon}}{l} \right)}.
\end{equation}
This entropy also satisfies the perimeter law if we identify
the invariant cutoff as
\begin{equation}
\label{invJ0}
\tilde{\epsilon}=l \sinh^{-1}\left(\frac{a}{l}\right)=l
\ln \left( \frac{a}{l}+\sqrt{1+ \left( \frac{a}{l} \right)^2}~~\right),
\end{equation}
where as expected $\tilde{\epsilon}$ is just a constant, and
independent of the mass and angular momentum of the black hole. This
feature appears to be an intrinsic property of horizon  
\cite{tho}.  Consequently, the entropy and invariant cutoff are all 
finite.

We now turn to the case of rotating black hole and deal with it
following the same steps adopted in the non-rotating case. The scalar
field is taken to be massless, $\mu =0$.
Since the black hole has now angular velocity, the modes of scalar
field are divided into two kinds of modes, which are superradiant (SR)
and non-superradiant (NSR) modes. The SR modes are the
common feature of rotating black holes like a Kerr black hole,
and are characterized by modes of
$0 \le E \le m\Omega_H$ and $m>0$ \cite{thor}, while the NSR 
modes are those of $E > m\Omega_H $ and any $m$. 
These two kinds of modes are distinct. Thus, we should divide the
range of energy integration in the expression of free energy
(\ref{free}) as 
\begin{equation}
\int dE = \int^\infty_{m\Omega_H} dE + \int^{m\Omega_H}_0 dE
\end{equation}
Then the free energy is divided into two parts: 
non-superradiant part $(F_{(NSR)})$
and superradiant part $(F_{(SR)})$,
\begin{equation}
\label{freeJ}
F_{(J\neq0)} = F_{(NSR)} + F_{(SR)},
\end{equation}
where we designate the free energy for the present case as
$F_{(J\neq0)}$ to distinguish it from that of the non-rotating case.

The NSR part of the free energy is given by
\begin{eqnarray}
F_{(NSR)} &=& -\frac{1}{\pi} \int dm \int^\infty_{m\Omega_H} dE
         \frac{1}{e^{\beta (E-m\Omega_H)} -1}
         \int^L_{r_H+\epsilon} dr k(r,m,E) \nonumber \\
\label{eq18}
    &=& -\frac{1}{\pi} \int dm \int^\infty_0 dE
         \frac{1}{e^{\beta E} -1} 
         \int^L_{r_H+\epsilon} dr k(r,m,E+m\Omega_H)~,
\end{eqnarray}
where at the second line changing of variable, $E \rightarrow
E+m\Omega_H$, was performed in order to make the integrations for $m$
and $E$ become independent. This expression for the NSR part is of the
same form as that of Eq. (\ref{free00}) except the details of the
radial wave number $k$. 
The integrations in Eq. (\ref{eq18}) can be 
evaluated straightforwardly and exactly.
Similar to the non-rotating case, the 
guideline for the evaluation of the physical 
free energy is to make the result of integration be real. 
Then the free energy is obtained as follows
\begin{eqnarray}
F_{(NSR)} &=& -\frac{1}{2} \int^\infty_0 dE
      \frac{1}{e^{\beta E}-1} \int^L_{r_H+\epsilon} dr 
       \frac{l E^2}{r (r_+^2-r_-^2) N^2(r)} 
     \frac{r_+^3 (r^2 - r^2_-)}{\sqrt{(r_+^2-r_-^2)(r^2-r_+^2)} }
           \nonumber \\
  &=& \frac{\zeta(3) l^3}{\beta^3}
       \left( \frac{r_+^2}{r_+^2-r_-^2} \right)^{3/2} 
       \left( \frac{1}{L} -
              \frac{1}{\sqrt{(r_H+\epsilon)^2-r_+^2} }
       \right)
             \nonumber \\
  &\stackrel{L \rightarrow \infty}{=}& 
       - \frac{\zeta(3) l^3}{\beta^3}
                   \left( \frac{r_+^2}{r_+^2-r_-^2} \right)^{3/2} 
                   \frac{1}{\sqrt{(r_H+\epsilon)^2-r_+^2} }~,
                  \label{fe}
\end{eqnarray}
where we have calculated the free energy for the nonextremal case so far.

On the other hand, the SR part of the free energy is given by
\begin{eqnarray}
F_{(SR)} &=& -\frac{1}{\pi} \int_{>0} dm \int^{m\Omega_H}_0 dE
        \frac{1}{e^{\beta (E-m\Omega_H)}-1} \int^L_{r_H+\epsilon} dr
        k(r,m,E)                   \nonumber \\
   &=& -\frac{\Omega_H}{\pi} \int_{>0} dm m^2 \int^1_0 dx
       \frac{1}{e^{-\beta m\Omega_H(1-x)}-1} \nonumber \\
   & & \hspace{0.5cm} \times \int^L_{r_H+\epsilon} dr
       \frac{1}{l r^2 N^2(r)}  
       \left[ ~ \frac{r_-^2}{r_+^2} (r^2 x - r_+^2)^2 
              -(r^2-r_+^2) (r^2-r_-^2)~
       \right]^{1/2}~,
                    \label{freeSR}
\end{eqnarray}
where the change of variable, $E \rightarrow x m\Omega_H$, was
performed at the second line. 
As in the previous evaluations, we first consider the $m$-integration.
In the present case, unlike the previous ones, we encounter a new
situation that each value of the azimuthal quantum number $m$ ranging 
from zero to infinity may contributes to $F_{(SR)}$. This means that 
for any value of $m$, the SR part of the free energy $F_{(J\neq0)}$ is 
real and thus physical, and there is no upper bound in the
$m$-integration. However, as we can see from Eq. (\ref{freeSR}),
the integration of the variable $m$ leads to the divergent result. A
certain factor, regulating factor, is thus needed to regulate the
$m$-integration.  Note that the $m$-integration appearing in the
calculations of $F_{(J=0)}$ and $F_{(NSR)}$ had bounded ranges of
integration variable and hence any regulating process was not needed.
We now take the regulating factor as $e^{-\epsilon_m m}$ and consider 
this in the $m$-integration. The evaluation of integration is
straightforward and the result is as follows.
\begin{equation}
\label{mint}
\int_0^\infty dm \frac{m^2~ e^{-\epsilon_m m}}{e^{-\beta \Omega_H (1-x) m}-1} 
= -\frac{2}{\epsilon_m^3}
  -\frac{2 \zeta(3)}{[\beta \Omega_H (1-x)]^3}~,
\end{equation}
which is obviously divergent as the regulating infinitesimal parameter
$\epsilon_m$ goes to zero, $\epsilon_m \rightarrow 0$. Next, what we
are going to do is the evaluation of the $x$-integration. The variable
$x$ ranges from zero to one. Due to the presence of the second term on
the right hand side of Eq. (\ref{mint}), the integrand entered in the
$x$-integration diverges at $x=1$, and thus the $x$-integration in
Eq. (\ref{freeSR}) becomes divergent. This divergence is related to that 
at $E=m\Omega_H$, since $x=1$ corresponds to $E=m\Omega_H$.
However, as we shall see later, this divergence does not appear in the
final expression of $F_{(SR)}$, which is obtained by requiring that it
is to be real. Since there is a divergence at $x=1$, we introduce a
new infinitesimal regulating cutoff $\epsilon_E$ in the
$x$-integration as follows: 
\begin{equation}
\int^{1-\epsilon_E}_0 dx~.
\end{equation}
With this prescription, we may evaluate the $x$-integration and expand
the result in powers of $\epsilon_E$. The expansion shows that there
are two divergent terms as follows:
\begin{equation}
\label{Ediv}
\frac{1}{2\epsilon^2_E} \sqrt{C}
-\frac{1}{\epsilon_E} \frac{r_-^2 r^2 (r^2-r_+^2)}{r_+^2 \sqrt{C}}~,
\end{equation}
where
\begin{equation}
\label{C}
C \equiv \frac{r^2}{r_+^2}(r_+^2-r_-^2) (r_+^2-r^2)~.
\end{equation}
The definition of $C$ shows that $C$ is less than zero
for $r>r_+$. This means that $\sqrt{C}$ is always imaginary for the
value of $r$ ranging from $r_++\epsilon$ to $L$. Thus divergent terms
in Eq. (\ref{Ediv}) do not contribute to the SR part of the physical
free energy, and only the terms of zeroth order in $\epsilon_E$ enter in
our consideration. Even among the terms of $\epsilon_E^0$-order, the
number of terms making $F_{(SR)}$ real is restricted. 
We now turn to the evaluation of the last $r$-integration. Unlike the
case of $F_{(NSR)}$ in Eq. (\ref{fe}), it may not be performed neatly
and compactly. However, since what is needed is the divergence
structure in terms of the brick wall cutoff $\epsilon$, we expand the
$r$-integration in powers of $\epsilon$ and evaluate only the terms
which diverge as $\epsilon \rightarrow 0$. Then the SR part of the
physical free energy is obtained as follows
\begin{eqnarray}
F_{(SR)} &=& \frac{l \Omega_H}{\pi} \int^{r_{\rm erg}}_{r_H+\epsilon} dr  
             \frac{1}{(r^2-r_+^2)(r^2-r_-^2)}    
       \left\{
      ~ \frac{r_+^2 \sqrt{r_{\rm erg}^2-r^2} }{\epsilon_m^3 r}
       -\frac{\zeta(3)}{(\beta \Omega_H)^3}
       \left[
            \frac{r_+^2 r \sqrt{r_{\rm erg}^2-r^2}}{r_+^2-r_-^2}
      \right. \right. 
        \nonumber \\  
  & & \hspace{2.5cm} \left. \left.
       + \frac{r_+r_-^2r (r^2-r_+^2)(r^2-r_-^2)}{[ (r_+^2-r_-^2)
                (r^2-r_+^2) ]^{3/2} }
       \tan^{-1} \left( \frac{r_+ \sqrt{r^2-r_+^2}}{ \sqrt{r_+^2-r_-^2} 
                              \sqrt{r_{\rm erg}^2-r^2} }
                 \right)~
      \right]~ 
      \right\}      
            \nonumber \\  
  &\approx& \frac{l \Omega_H}{\pi} \frac{r_-}{r_+^2-r_-^2} 
            \left[ \frac{1}{2 \epsilon_m^3} - 
                   \frac{\zeta (3)}{ (\beta \Omega_H)^3}
                   \frac{r_+^2}{r_+^2-r_-^2}
            \right]        
            \ln \left( \frac{r_+}{\epsilon} \right) 
            + {\cal O}(\epsilon^0)~.
         \label{finalSR}
\end{eqnarray}
The maximum value of $r$ in the $r$-integration changed to
$r_{\rm erg}$ due to the result of the $x$-integration. This implies
that only the SR modes in the radial region between the brick wall and 
stationary limit, i.e., ergosphere, contribute to the SR part of
the physical free energy. Comparing with the NSR part $F_{(NSR)}$ of
the free energy in Eq. (\ref{fe}), $F_{(SR)}$ in (\ref{finalSR})
shows two different aspects. First is the appearance of the divergent
structure, $1/\epsilon^3_m$, due to the large azimuthal quantum
number, which is absent in $F_{(NSR)}$. Second is that the leading
order divergence in terms of the brick wall cutoff $\epsilon$ is
logarithmic ($\ln \epsilon$), while it is power like
($1/\sqrt{\epsilon}$) in the case of $F_{(NSR)}$.
Thus it may be concluded that the SR part gives the sub-leading order
contribution to the free energy $F_{(J\neq0)}$, while the NSR part
gives the leading order one. 

Before considering the entropy for the case of rotating black
hole, we would like to comment on the non-rotating limit of
Eq. (\ref{finalSR}). From the expression of $F_{(SR)}$ in
Eq. (\ref{freeSR}), it is obvious that $F_{(SR)}$ is zero when the 
limit $\Omega_H \rightarrow 0$ is taken. It is natural and implies
that the SR modes do not exist in the non-rotating case, as it should
be. However, $F_{(SR)}$ in Eq. (\ref{finalSR}) diverges under this
limit. This inconsistency may be explained as follows: up to now, we
have given a formulation retaining $\Omega_H$ as a non-zero finite
value, or more precisely $r_-\neq0$. That $\Omega_H$ is not zero means
that the ergosphere does exist and hence $r_{\rm erg} > r_H$.  If we
now assume that $\Omega_H$ is so small such that $r_{\rm erg} <
r_H+\epsilon$, it happens that we count the SR modes inside the brick
wall, as can be known from the range of $r$-integration in
Eq. (\ref{finalSR}). But, the essence of the 
brick wall method is to count the modes 
outside the brick wall. Therefore, in the framework of the method, the
evaluation of $F_{(SR)}$ for the black hole rotating with angular
velocity lower than a certain value becomes totally meaningless. 
The condition which $\Omega_H$ has to satisfy may now be determined
by looking at the inequality, $r_{\rm erg} > r_H+\epsilon$. After a
short manipulation by using Eqs. (\ref{erg}) and (\ref{angular}), the 
condition is obtained as follows:
\begin{equation}
\label{avel}
\Omega_H^2 > \frac{1}{l^2} 
	\left[ 
		\left( 1+\frac{\epsilon}{r_+} \right)^2 -1
	\right]~.
\end{equation}
This may be also understood as the validity condition for the final
expression of $F_{(SR)}$ in Eq. (\ref{finalSR}). The argument given up 
to now explains why we cope with the non-rotating and the rotating
black holes individually.

We are now ready to evaluate the entropy for the massless field as for
the case of the rotating black hole. The physical free energy
$F_{(J\neq0)}$ is obtained by substituting the results of Eqs. (\ref{fe})
and (\ref{finalSR}) into Eq. (\ref{freeJ}):
\begin{equation}
F_{(J\neq0)} \approx
           - \frac{l^3 \zeta(3)}{\sqrt{2} \beta^3} 
             \left( \frac{r_+}{r_+^2-r_-^2} \right)^{3/2}
             \frac{r_+}{\sqrt{\epsilon}} 
          +  \frac{l \Omega_H}{\pi} \frac{r_-}{r_+^2-r_-^2} 
            \left[ \frac{1}{2 \epsilon_m^3} - 
                   \frac{\zeta (3)}{ (\beta \Omega_H)^3}
                   \frac{r_+^2}{r_+^2-r_-^2}
            \right]        
            \ln \left( \frac{r_+}{\epsilon} \right)~,
\label{FJ}
\end{equation}
where the NSR part $F_{(NSR)}$ is expanded in powers of $\epsilon$ as
$F_{(SR)}$ was. Unlike the free energy $F_{(J=0)}$ for the case of
non-rotating black hole,  $F_{(J\neq0)}$ has an additional divergence
due to the large azimuthal quantum number, which comes from the SR
modes, besides that due to the brick wall cutoff. 
This additional divergence may be problematic when we consider the
relation between the entropy and the brick wall cutoff. Fortunately,
however, it does not matter because the entropy is obtained by
differentiating the free energy by the inverse temperature
$\beta$, but the term related to the divergence $1/\epsilon_m^3$ in
Eq. (\ref{FJ}) does not depend on $\beta$. Therefore, the entropy is
safe from the additional divergence and, from the standard formula, is
obtained as follows:
\begin{eqnarray}
S_{(J\neq0)} &=& \beta^2 
                 \left. 
                   \frac{\partial F_{(J\neq0)}}{\partial \beta}
                 \right|_{\beta=\beta_H} 
        			\nonumber \\
  &\approx& \frac{ 3 \zeta (3)}{4\pi^2 l}
            \left[ \sqrt{ \frac{ r_+(r_+^2-r_-^2) }{2} }
                   \frac{1}{\sqrt{\epsilon}}
                  + \frac{r_+^2}{\pi r_-} \ln \left( \frac{r_+}{\epsilon}
                                              \right)
            \right]~,
  \label{entJ}
\end{eqnarray}
where $\beta_H$ is for the rotating black hole and is given by 
\begin{equation}
\beta_H = \frac{2\pi r_+ l^2}{r_+^2 - r_-^2}~.
\end{equation}
We now take only the leading order term in terms of $\epsilon$ in the
r.h.s. of Eq. (\ref{entJ}), which is the contribution from the NSR
modes. Obviously, there is a contribution from the SR modes, which is 
however sub-leading and hence may be neglected. 
We now determine the value of $\epsilon$ by equating
$S_{(J\neq0)}$ with the area (perimeter) law (\ref{area}).
Then it is given by
\begin{equation}
\epsilon \approx \frac{r_+}{2} \left( \frac{a}{l} \right)^2
                 \left( 1 - \frac{r_-^2}{r_+^2} \right)~,
\end{equation}
where $a$ is the number defined in Eq. (\ref{adef}).
The invariant distance from the horizon to the brick wall is
calculated by using Eq. (\ref{invariant}) and is obtained as follows:
\begin{equation}
\label{invJ}
\tilde{\epsilon} = a + {\cal O} (a^2)~,
\end{equation}
which is independent on the mass and angular velocity of the rotating
black hole, as was in the case of the non-rotating black hole.
It should be noted here that if the
invariant distance of Eq. (\ref{invJ0}) is expanded in powers of $a$
then it is just the same with Eq. (\ref{invJ}) at the leading
order.  
The entropy may now be rewritten in terms of $\tilde{\epsilon}$ as
\begin{equation}
S_{(J\neq0)} \approx 4 \pi a \frac{r_+}{\tilde{\epsilon}} ~,
\end{equation}
which satisfies the perimeter law if we take Eq. (\ref{invJ}) as the
expression for $\tilde{\epsilon}$. Since the entropy for the
non-rotating case also satisfies the perimeter law, it may be
concluded that the perimeter law is always satisfied regardless of
whether or not the black hole is rotating, at least in the framework 
of the brick wall method. 

Before concluding remark, we would like to compare our results with
those of ref. \cite{ich} where the authors investigate the
thermodynamics of scalar fields in the BTZ black hole backgrounds and
conclude that generically the entropy does not satisfy the perimeter
law. As in our formulation, there appears three regularization cutoffs
in ref. \cite{ich}:  
an infinitesimal distance from the horizon ($\epsilon_H$),
cutoff for the occupation number of the particle ($N_1$) for each mode
satisfying $E-m\Omega_H \leq 0$, 
and cutoff for the absolute value of quantum number ($N_2$).
Comparison between these cutoffs and those of us shows that
$\epsilon_H$ is related to the brick wall cutoff $\epsilon$, $N_1$
corresponds to $\epsilon_E$ related to the divergence at
$E=m\Omega_H$, and  $N_2$ corresponds to the large azimuthal quantum
number which leads to the introduction of the infinitesimal parameter
$\epsilon_m$. These three cutoffs remain in the final expression for
the entropy, which may give difficulty in deriving the perimeter law
by adjusting the cutoffs.  However, as for our case, only the brick
wall cutoff $\epsilon$ appears in the entropy while the other two
parameters disappear during the formulation by natural and physical
reasons, and the perimeter law is obtained by adjusting $\epsilon$.  

In conclusion, we have studied the entropy of a spinless and massless
scalar field around the non-extremal BTZ black hole,
which is similar to the Kerr metric but not asymptotically flat.
The area law has been obtained by adjusting the location of the brick
wall, which is just outside the horizon for both the cases of
non-rotating and rotating black hole. As for the entropy in the
rotating case, there was the contribution from the superradiant modes,
which was however sub-leading, while the non-superradiant modes gave
the leading order contribution. Although there have appeared three
types of regulating parameters during our formulation, all of them
except the brick wall cutoff $\epsilon$ were absent at the final
expression of the entropy. It has been shown that, for the
consistency of the brick wall method, the angular velocity of the
rotating black hole should be greater than a certain value represented
by Eq. (\ref{avel}). This led us to consider the rotating and the
non-rotating black holes separately.

\section*{Acknowledgments}
S.-W. Kim and W. T. Kim were partially supported by
Korea Science and Engineering Foundation, 95-0702-04-01-3.
H. Shin and Y.-J. Park were supported in part
by Ministry of Education, Project No. BSRI-95-2414.
W. T. Kim was supported in part by the Korea Science and
Engineering Foundation through the Center for Theoretical
Physics(1996).


\end{document}